\documentclass[aps,preprintnumbers, superscriptaddress, amsmath, showpacs]{revtex4}
\usepackage{epsfig}
\usepackage{psfrag}

\usepackage{graphicx}
\usepackage{dcolumn}
\usepackage{bm}

\begin{document}

\preprint{RU06-9-B}

\title{Angular Momentum Mixing in Single Flavor Color Superconductivity with Transverse Pairing}

\author{Bo  Feng}
 \email{fengbo@iopp. ccnu. edu. cn}
\affiliation{Institute of Particle Physics,  Huazhong Normal University,
Wuhan,  430079,  China}
\author{ De-fu  Hou}%
 \email{hdf@iopp. ccnu. edu. cn}
 \affiliation{Institute of Particle Physics,  Huazhong Normal University,
Wuhan,  430079,  China}
\author{Hai-cang Ren}
\email{ren@mail.rockefeller.edu} \affiliation{Physics Department, The
Rockefeller University, 1230 York Avenue,  New York,  NY
10021-6399} \affiliation{Institute of Particle Physics, Huazhong
Normal University, Wuhan,  430079,  China}

\date{\today}

\begin{abstract}
Because of the equal strength of the pairing potential mediated by
one-gluon exchange for all partial waves to the leading order QCD
running coupling constant and the nonlinearity of the gap
equation, the non-spherical pairing in single flavor color
superconductivity(CSC) can not be restricted in a single
non-$s$-wave channel and the mixing among different angular
momenta will occur. In this paper, we examine the angular momentum
mixing in single flavor CSC with transverse pairing, in which the
pairing quarks have opposite helicity. We find that the free
energy of all non-spherical pairing states are lowered by angular
momentum mixing compared with that contain p-wave only. But the
amount of the free energy drop is numerically small. Consequently
the most stable pairing state that respect the time reversal
invariance remains the spherical CSL.
\end{abstract}

\maketitle

\section{Introduction}\label{sec:level1}
Color superconductivity(CSC) is expected to be the ground state in
a quark matter of sufficiently high baryon density but low
temperature \cite{B,DA,ARW1,RTEM,MKF,dk,MAKT}. The most favorable
pairing channel in the chiral limit is between quarks of different
flavors. At the baryon density accessible in a compact star,
however, the nonzero strange quark mass induces a significant
mismatch between Fermi momenta of different flavors that
suppresses the phase space available for pairing. A lot of exotic
CSC states or mixed states have been proposed in the literature
\cite{HZC,MI,MCK,ABR,BCR}. But the result of their competition
remains under debate.

Since the single flavor pairing is free from the Fermi momentum
mismatch, it has been considered as an interesting alternative in
this circumstance. The corresponding gap function in this case
should contain higher total angular momentum as required by Pauli
principle, the obvious alternative is to consider the p-wave
pairing($J=1$)\cite{T,A,SSW,SWR,D,RD}. A large variety of pairing
structures of the single flavor CSC has been discussed in the
literature\cite{T,A}. Except for the color-spin-locked(CSL) state,
none of the energy gaps is spherical. All these works restrict the
angular dependence of a non-spherical gap within a single partial
wave, $p$-wave, which is valid if the pairing potential consists
of the $p$-channel only, such as that in the superfluid ${}^3He$.
The forward singularity of diquark scattering in QCD, however,
brings in all angular momentum channels to the pairing potential.
Because of the nonlinearity of the gap equation, all partial waves
contribute to a non-spherical energy gap at the {\it {same}} order
of the QCD running coupling. In a previous paper\cite{BDH}, we
examined this angular momentum mixing effect in the non-spherical
CSC with longitudinal pairing, in which the two pairing quarks
carry the same helicity. We found that the energy gap ( gap
function evaluated at zero Matsubara energy ) takes the form
\begin{equation}
\phi(0,\hat{\bf p})=\frac{2048\sqrt{2}\pi^4\mu}{N_f^{5/2}g^5}
e^{-\frac{3\pi^2}{\sqrt{2}g}-\frac{\pi^2+4}{8}}f(\hat{\bf p})
\end{equation}
where $\mu$ is the chemical potential, $\hat{\bf p}$ is the
direction of the relative momentum in a Cooper pair and the
function $f(\hat{\bf p})$ is an infinite series in spherical
harmonics of all $J$'s. In particular, we find
\begin{equation}
f(\hat{\bf
p})=e^{1/3}e^{-6}[0.9866P_1(\cos\theta)-0.0673P_3(\cos\theta)+0.0214P_5(\cos\theta)+...]
\end{equation}
in contrast with $f(\hat{\bf p})=e^{1/3}e^{-6}P_1(\cos\theta)$
reported in the literature for the polar phase. The contributions from higher partial waves,
however, are numerically small. The same effect is
expected for the transverse pairing and the parallel analysis will
be carried out in this paper.

The free energy density of the superconductivity with angular
momentum mixing will be lowered relative to that containing p-wave
pairing only. In the toy model considered in \cite{BDH}, which
maximizes the forward singularity, the magnitude of the condensation energy is
increased by a factor of 1.54. Without including the angular
momentum mixing, the gaining factor of the magnitude of the condensation energy in
CSC by forming the spherical CSL state compared with the
non-spherical polar state is 1.54 for the longitudinal pairing and
1.14 for the transverse pairing \cite{A}. This raises the issue to
recalculate the condensation energy of various non-spherical
pairing states of the single-flavor CSC in the light of the
angular momentum mixing and to compare the result with that of the spherical CSL.
So we did in previous works \cite{BDH}
\cite{BDH1} for the polar and A phases of the longitudinal
pairing. The formulation we developed can be carried over readily
to the polar and A phases of the transverse pairing and the
parallel analysis is made in this work. We found that, like the
longitudinal pairing case, the amount of mixing is numerically
small so the slightly reduced free energy of the polar and A
phases is not low enough to compete with the spherical CSL phase.

This paper is organized as follows. In Sec. II, we examine the
quark propagator with transverse pairing. The condensation energy
density and the gap equation will be obtained from the CJT
effective action in Section III.  The gap equation will be reduced
to an nonlinear integral equation for the angular dependence with
zero and nonzero azimuthal quantum number in Section IV and their
numerical solutions will be given there as well. We conclude in
Sec. V. Some technical details about the full quark propagators
and the relevant spherical harmonics are deferred to Appendices A
and B. Our convention for the metric tensor is $g_{\mu\nu}={\rm
diag}\{1,-1,-1,-1\}$. Our units are $\hbar=c=k_B=1$ and 4-vectors
are denoted by capital letters, $K \equiv K^\mu = (i\nu, {\bf k})$
with $\nu$ the Matsubara energy, which becomes continuous at zero
temperature, $T=0$.

\section{The quark propagator for transverse pairing}

Though the general structure of the quark propagator in CSC state
has been discussed extensively in the literature \cite{DR}, the formulation
we developed below is particularly suitable to explore the angular
momentum mixing in a transverse pairing state.

We start with the thermal Nambu-Gorkov quark propagator ${\cal S}(X)$, defined as
\begin{equation}
S_{\alpha\beta}(X)=<{\cal T}[\Psi_\alpha(X)\bar\Psi_\beta(0)]>
=T\sum_\nu\int \frac{d^3{\bf p}}{(2\pi)^3}e^{iP\cdot X}{\cal S}(P)
\label{definition}
\end{equation}
where $\Psi$ and $\bar \Psi$ are the Nambu-Gorkov(NG) spinors,
$P=(i\nu,{\bf p})$, $P\cdot X=-\nu\tau-\bf p\cdot\bf x$ with $\nu$
the Matsubara energy and $\tau$ the Euclidean time, ${\cal T}$
imposes the Euclidean time ordering and ($\alpha$, $\beta$) run
over the NG, Dirac and color indexes. At zero temperature,
$T\sum_\nu=\int_{-\infty}^\infty\frac{d\nu}{2\pi}$. The NG spinors
are defined by
\begin{equation}
\Psi= \left( \begin{array}{c} \psi\\
\psi_C
\end{array}
\right), \hspace{0.3cm}\bar \Psi=(\bar \psi, \bar \psi_C),
\label{2}
\end{equation}
where, $\psi$ and $\bar\psi$ are written in terms of the ordinary
quark spinors, $\psi_C=C\tilde{{\bar \psi}}$ and $\bar\psi_C=\tilde\psi C$
are the charge-conjugate spinors with the tilde standing for the transpose,
and $C\equiv i\gamma^2\gamma^0$. It follows from the
definition (\ref{definition}) and the self-conjugation relation
$\Psi={\cal C}\Psi^*$ with
\begin{equation}
{\cal C}=\left( \begin{array}{cc}
   0 & {i\gamma_2}  \\
   {i\gamma_2} & {0}
\end{array} \right).
\end{equation}
that
\begin{equation}
S^*(P)={\cal C}^\dagger S(\bar P){\cal C}
\label{conjugation}
\end{equation}
with $\bar{P}=(i\nu,-{\bf p})$ and
\begin{equation}
\tilde{\cal S}(P)=\Gamma_0{\cal C}{\cal S}(-P)\Gamma_0{\cal C}
\label{antisymm}
\end{equation}
with
\begin{equation}
\Gamma_0=\left( \begin{array}{cc}
   {\gamma_0} & {0}  \\
   {0} & {\gamma_0}
\end{array} \right).
\end{equation}
The relation (\ref{conjugation}) was proved in \cite{BDJH} and the proof of
(\ref{antisymm}) proceeds similarly. The NG structure of the
inverse propagator is
\begin{equation}
S^{-1}(P)\equiv \left( \begin{array}{cc}
   {K^+(P)} & {\Phi^-(P)}  \\
   {\Phi^+(P)} & {K^-(P) }
\end{array} \right).
\label{3}
\end{equation}
where each block is a $4\times4$ matrix with respect to
Dirac indexes and a $3\times 3$ matrix in color indexes. We shall focus on the
off-diagonal blocks $\Phi^{\pm}(P)$ below.

It is convenient to work in the chiral representation where
\begin{equation}
\psi=\left ( \begin{array}{c} \chi_R\\
\chi_L
\end{array}
 \right), \hspace{0.3cm}\bar \psi=(\chi^+_L, \chi^+_R),
\label{5}
\end{equation}
and their charge-conjugate counterparts
\begin{equation}
\psi_C=\left ( \begin{array}{c} i\sigma^2\chi^*_L\\
-i\sigma^2\chi^*_R
\end{array}
 \right), \hspace{0.3cm}\bar{\psi}_C=(i\tilde\chi_R\sigma^2, -i\tilde{\chi}_L\sigma^2),
\label{6}
\end{equation}
with $\chi_L$($\chi_R$) the two-component left-handed(right-handed spinors)
and $\sigma^2$ the second Pauli matrix. The matrix
$\Phi^\pm(P)$ are decomposed accordingly. For example,
\begin{equation}
\Phi^-(P)\equiv \left( \begin{array}{cc}
   {\Phi^-_{++}(P)} & {\Phi^-_{+-}(P)}  \\
   {\Phi^-_{-+}(P)} & {\Phi^-_{--}(P) }
\end{array} \right).
\label{7}
\end{equation}
with each block a $2\times 2$ matrix with respect to spinor
indexes. For transverse pairing between different chiralities,
$\Phi^\pm_{++}(P)=\Phi^\pm_{--}(P)=0$. Next, we consider the
off-diagonal block $\Phi^-_{+-}(P)$ and write
\begin{equation}
\Phi^-_{+-}(P)=\varphi(P)+\varphi_0(P){\bm\sigma}\cdot\hat{\bf p}
+\varphi_+(P){\bm\sigma}\cdot\hat{\bf e}_+(\hat{\bf p})
+\varphi_-(P){\bm\sigma}\cdot\hat{\bf e}_-(\hat{\bf p}),
\end{equation}
where the circular basis $\hat{\bf e}_{\pm}(\hat{\bf
p})=\frac{1}{\sqrt{2}}(\hat{\bm\theta}\pm i\hat{\bm\varphi})$ with
$\hat{\bm\theta}$ and $\hat{\bm\varphi}$ the unit vectors of the
spherical coordinates of $\bf p$. We have
\begin{eqnarray}
\hat{\bf p} &=& (\sin\theta\cos\varphi,\sin\theta\sin\varphi,\cos\theta)\nonumber\\
\hat{\bm\theta} &=& (\cos\theta\cos\varphi,\cos\theta\sin\varphi,-\sin\theta)\nonumber\\
\hat{\bm\varphi} &=& (-\sin\varphi,\cos\varphi,0),
\end{eqnarray}
\begin{equation}
\hat{\bf e}_+(-\hat{\bf p})=\hat{\bf e}_-(\hat{\bf p}) \qquad
\hat{\bf e}_+(\hat{\bf p})^*=\hat{\bf e}_-(\hat{\bf p}),
\end{equation}
and
\begin{equation}
\hat{\bf p}\times\hat{\bf e}_\pm(\hat{\bf p})=\mp i\hat{\bf e}_\pm(\hat{\bf p}).
\label{eigenvalue}
\end{equation}

Performing a rotation about $\bf p$ by an angle $\gamma$, we find
\begin{equation}
\Phi^-_{+-}(P)\to e^{-\frac{i}{2}\gamma{\bm\sigma}\cdot\hat{\bf
p}}\Phi^-_{+-}(P)
 e^{\frac{i}{2}\gamma{\bm\sigma}\cdot\hat{\bf p}}
=\varphi(P)+\varphi_0(P){\bm\sigma}\cdot\hat{\bf p}
+e^{-i\gamma}\varphi_+(P){\bm\sigma}\cdot\hat{\bf e}_+(\hat{\bf p})
+e^{i\gamma}\varphi_-(P){\bm\sigma}\cdot\hat{\bf e}_-(\hat{\bf p}).
\end{equation}
For a Copper pair of two quarks or two antiquarks of opposite
helicity, the projection of its angular momentum in the direction
$\bf p$ has to be $\pm 1$, which implies that
$\varphi(P)=\varphi_0(P)=0$. Up to now, the coefficients $\varphi$'s
remain $3\times 3$ antisymmetric matrices in color space. For the
polar phase and A phase considered in this work, we set
\begin{equation}
\Phi^-_{+-}(P)={\bm\sigma}\cdot{\bm\Psi}(P)\lambda_2
\end{equation}
where
\begin{equation}
{\bm\Psi}(P)=\phi_+(P)\hat{\bf e}_+(\hat{\bf p})+\phi_-(P)\hat{\bf
e}_-(\hat{\bf p}). \label{circular}
\end{equation}
and $\lambda_2$ is the second Gell-Mann matrix.

The Dirac structure of other off-diagonal elements in Eq.(\ref{3})
can be deduced from the relations (\ref{conjugation}) and
(\ref{antisymm}). We now could write down the inverse full quark
propagator in chiral representation using the Dyson-Schwinger
equation and ignoring the contribution from the wave-function
renormalization, and reads
\begin{equation}
S^{-1}(P)= \left(\begin{array}{cccc} 0 &
{i\nu+\mu-{\bm\sigma}\cdot{\bf p}} & 0 &
{\lambda_2{\bm\sigma}\cdot{\bm\Psi(P)}}\\
{i\nu+\mu+{\bm\sigma}\cdot{\bf p}} & 0 &
{-\lambda_2{\bm\sigma}\cdot{\bm\Psi(-P)}} & 0 \\
0 & {\lambda_2{\bm\sigma}\cdot{\bm{\Psi}}^*(\bar P)} & 0 &
{i\nu-\mu-{\bm\sigma}\cdot{\bf p}}\\
{-\lambda_2{\bm\sigma}\cdot{\bm\Psi}^*(-\bar P)} & 0 &
{i\nu-\mu+{\bm\sigma}\cdot{\bf p}} & 0
\end{array} \right )
\label{21}
\end{equation}
Inverting the matrix (\ref{21})(see the appendix A for details),
we find that
\begin{equation}
S(P)=\left( \begin{array}{cccc} S_{11} &
S_{12} & S_{13} & S_{14}\\ S_{21} & S_{22} & S_{23} & S_{24}\\
S_{31} & S_{32} & S_{33} & S_{34}\\ S_{41} & S_{42} & S_{43} &
S_{44}
\end{array}\right)\label{blockes}
\end{equation}
with
\begin{subequations}
\begin{eqnarray}
S_{11}&=&S_{13}=S_{22}=S_{24}=S_{31}=S_{33}=S_{42}=S_{44}=0\\
S_{12}&=&-\frac{i\nu-(p+\mu)}{\nu^2+(p+\mu)^2+2\lambda_2^2\phi_-(-P)\phi_-^*(-\bar
P)}\frac{1+{\bm\sigma}\cdot\hat{\bf
p}}{2}-\frac{i\nu+(p-\mu)}{\nu^2+(p-\mu)^2+2\lambda_2^2\phi_+(-P)\phi_+^*(-\bar
P)}\frac{1-{\bm\sigma}\cdot\hat{\bf p}}{2}\\
S_{14} &=&-\frac{\lambda_2{\bm\sigma}\cdot\hat{\bf e}_+(\hat{\bf
p})\phi_-(-P)} {\nu^2+(p+\mu)^2+2\lambda_2^2\phi_-(-P)\phi_-^*(-\bar
P)}\frac{1-{\bm\sigma}\cdot\hat{\bf
p}}{2}-\frac{\lambda_2{\bm\sigma}\cdot\hat{\bf e}_-(\hat{\bf
p})\phi_+(-P)} {\nu^2+(p-\mu)^2+2\lambda_2^2\phi_+(-P)\phi_+^*(-\bar
P)}\frac{1+{\bm\sigma}\cdot\hat{\bf p}}{2}\\
S_{21}&=&-\frac{i\nu-(p+\mu)}{\nu^2+(p+\mu)^2+2\lambda_2^2\phi_-(P)\phi_-^*(\bar
P)}\frac{1-{\bm\sigma}\cdot\hat{\bf
p}}{2}-\frac{i\nu+(p-\mu)}{\nu^2+(p-\mu)^2+2\lambda_2^2\phi_+(P)\phi_+^*(\bar
P)}\frac{1+{\bm\sigma}\cdot\hat{\bf p}}{2}\\
S_{23}&=&\frac{\lambda_2{\bm\sigma}\cdot\hat{\bf e}_-(\hat{\bf
p})\phi_-(P)} {\nu^2+(p+\mu)^2+2\lambda_2^2\phi_-(P)\phi_-^*(\bar
P)}\frac{1+{\bm\sigma}\cdot\hat{\bf
p}}{2}+\frac{\lambda_2{\bm\sigma}\cdot\hat{\bf e}_+(\hat{\bf
p})\phi_+(P)} {\nu^2+(p-\mu)^2+2\lambda_2^2\phi_+(P)\phi_+^*(\bar
P)}\frac{1-{\bm\sigma}\cdot\hat{\bf p}}{2}\\
S_{32} &=& -\frac{\lambda_2{\bm\sigma}\cdot\hat{\bf e}_-(\hat{\bf
p})\phi_-^*(-\bar P)}
{\nu^2+(p+\mu)^2+2\lambda_2^2\phi_-(-P)\phi_-^*(-\bar
P)}\frac{1+{\bm\sigma}\cdot\hat{\bf
p}}{2}-\frac{\lambda_2{\bm\sigma}\cdot\hat{\bf e}_+(\hat{\bf
p})\phi_+^*(-\bar P)}
{\nu^2+(p-\mu)^2+2\lambda_2^2\phi_+(-P)\phi_+^*(-\bar
P)}\frac{1-{\bm\sigma}\cdot\hat{\bf p}}{2}\\
S_{34}&=&-\frac{i\nu+(p+\mu)}{\nu^2+(p+\mu)^2+2\lambda_2^2\phi_-(-P)\phi_-^*(-\bar
P)}\frac{1-{\bm\sigma}\cdot\hat{\bf
p}}{2}-\frac{i\nu-(p-\mu)}{\nu^2+(p-\mu)^2+2\lambda_2^2\phi_+(-P)\phi_+^*(-\bar
P)}\frac{1+{\bm\sigma}\cdot\hat{\bf p}}{2}\\
S_{41}&=&\frac{\lambda_2{\bm\sigma}\cdot\hat{\bf e}_+(\hat{\bf
p})\phi_-^*(\bar P)}
{\nu^2+(p+\mu)^2+2\lambda_2^2\phi_-(P)\phi_-^*(\bar
P)}\frac{1-{\bm\sigma}\cdot\hat{\bf
p}}{2}+\frac{\lambda_2{\bm\sigma}\cdot\hat{\bf e}_-(\hat{\bf
p})\phi_+^*(\bar P)}
{\nu^2+(p-\mu)^2+2\lambda_2^2\phi_+(P)\phi_+^*(\bar
P)}\frac{1+{\bm\sigma}\cdot\hat{\bf p}}{2}\\
S_{43}&=&-\frac{i\nu+(p+\mu)}{\nu^2+(p+\mu)^2+2\lambda_2^2\phi_-(P)\phi_-^*(\bar
P)}\frac{1+{\bm\sigma}\cdot\hat{\bf
p}}{2}-\frac{i\nu-(p-\mu)}{\nu^2+(p-\mu)^2+2\lambda_2^2\phi_+(P)\phi_+^*(\bar
P)}\frac{1-{\bm\sigma}\cdot\hat{\bf p}}{2}.
\end{eqnarray}\label{25}
\end{subequations}
We notice that $\phi_+$ associates with particles and $\phi_-$ with antiparticles.
The is because we designate implicitly the momentum ${\bf p}$ to the
quark of positive helicity in a Cooper pair. Since the contribution from
the Cooper pairs of antiquarks goes beyond the subleading order, we shall
approximate $\phi_-(P)\simeq 0$ and introduce the gap function $\phi(\nu,{\bf p})$ via
\begin{equation}
\phi_+(P)\equiv\frac{1}{\sqrt{2}}\phi(\nu,{\bf p}).
\label{gapfunction}
\end{equation}
As is the same with 2SC, the solution to the gap equation gives
rise to an even function in the Matsubara energy, we have
$\phi_+(-P)=\frac{1}{\sqrt{2}}\phi(\nu,-{\bf p})$, $\phi_+(\bar
P)=\frac{1}{\sqrt{2}}\phi(\nu,{\bf p})$ and $\phi_+(-\bar
P)=\frac{1}{\sqrt{2}}\phi(\nu,-{\bf p})$.

\section{The gap equation}
In this section, we shall derive the gap equation for the transverse
pairing. The starting point is the CJT free energy, which can be
approximated in weak coupling as\cite{dk,CJT,BDH}
\begin{equation}
\Gamma = \Gamma_n + \Omega F
\end{equation}
where, $\Gamma_n$ is the free energy of the normal phase
and $F$ the condensation energy density, which is the part of $\Gamma$
responsible to the gap equation. The CJT formula for $F$ reads
\begin{equation}
F=-\frac{1}{2\Omega}[{\rm Trln}D^{-1}-{\rm Trln}D_n^{-1}-{\rm
Trln}S^{-1} +{\rm Trln}S_0^{-1}-{\rm Tr}(S_{0}^{-1}S-1)]
\label{28}
\end{equation}
where $D$ and $S$ are the full propagators for gluon and quark
respectively, $S_0$ is the tree level quark propagator and $D_n$
is the hard-dense-loop (HDL) gluon propagator. Following the
procedure of \cite{IDHD,JHIDD}, we approximate
\begin{equation}
{\rm Trln}D^{-1}-{\rm Trln}[D_n^{-1}]\simeq{\rm Tr}[D_n\delta\Pi]
\label{HF}
\end{equation}
where
\begin{equation}
\delta\Pi=\Pi-\Pi_n,
\end{equation}
with $\Pi_n$ the HDL resummed gluon self-energy in normal phase.
Eq.(\ref{HF}) represents the Hartree-Fock energy of the mean-field theory.
In the Coulomb gauge, the HDL gluon propagator is
\begin{equation}
D_{n,00}(K)=D_l(K), \hspace{0.3cm}
D_{n,0i}(K)=D_{n,i0}(K)=0,\hspace{0.3cm}
D_{n,ij}(K)=(\delta_{ij}-\hat{k}_i\hat{k}_j)D_t(K) \label{18}
\end{equation}
where $D_{l,t}$ are the longitudinal and transverse propagators
respectively and are diagonal in adjoint color space, i.e.
$D_{l,t}^{ab}=\delta^{ab} D_{l,t}$. Consequently, we only need the
00-component, $\Pi^{00}(K)$, and the transverse projection of the
ij-components,
\begin{equation}
(\delta_{ij}-\hat{k}_i\hat{k}_j)\Pi^{ij}(K)=\Pi^{ii}(K)-\hat{k}_i\hat{k}_j\Pi^{ij}(K)
\end{equation}
The gluon self-energy in super phase reads
\begin{equation}
\Pi_{ab}^{\mu\nu}(K)=\frac{1}{2}g^2\sum\limits_{P}{\rm
Tr}\Big{[}\hat{\Gamma}_{a}^{\mu}S(P)\hat{\Gamma}_{b}^{\nu}S(P^{\prime})\Big{]}
\label{gsenery}
\end{equation} where $P^{\prime}=P-K$, $\sum_P\equiv T\sum_{\nu}\int\frac{d^3{\bf p}}{(2\pi)^3}$ and
\begin{equation}
\hat\Gamma_{a}^{\mu}\equiv \left (
\begin{array}{cccc}
0 & \sigma^{\mu}T_a & 0 &0 \\
\bar{\sigma}^{\mu}T_a & 0 & 0 & 0 \\
0 & 0 & 0 & -\sigma^{\mu}\tilde T_a \\
0 & 0 & -\bar{\sigma}^{\mu}\tilde T_a & 0
\end{array} \right )
\label{vertex}
\end{equation}
with $\sigma^{\mu}\equiv(1,{\bm\sigma})$,
$\bar{\sigma}^{\mu}\equiv(1,-{\bm\sigma})$, and $T_a=\lambda_a/2 $
$(a=1,...,8)$ is the $a$th $SU(3)$ generator with $\lambda_a$ the
$a$th Gell-Mann matrix. Substituting Eq.(\ref{blockes}) and
(\ref{vertex}) into Eq.(\ref{gsenery}), we have
\begin{eqnarray}
\nonumber
\Pi_{ab}^{\mu\nu}(K)=\frac{1}{2}g^2\sum\limits_{P}{\rm
Tr}\Big{[}\sigma^{\mu}T_aS_{21}(P)\sigma^{\nu}T_bS_{21}(P^{\prime})+\bar{\sigma}^{\mu}T_aS_{12}(P)\bar{\sigma}^{\nu}T_bS_{12}(P^{\prime})\\
\nonumber
-\sigma^{\mu}T_aS_{23}(P)\sigma^{\nu}\tilde T_bS_{41}(P^{\prime})-\bar{\sigma}^{\mu}T_aS_{14}(P)\bar{\sigma}^{\nu}\tilde T_bS_{32}(P^{\prime})\\
\nonumber\hfill \\
\nonumber
-\sigma^{\mu}\tilde T_aS_{41}(P)\sigma^{\nu}T_bS_{23}(P^{\prime})-\bar{\sigma}^{\mu}\tilde T_aS_{32}(P)\bar{\sigma}^{\nu}T_bS_{14}(P^{\prime})\\
\nonumber\hfill\\+\sigma^{\mu}\tilde
T_aS_{43}(P)\sigma^{\nu}\tilde
T_bS_{43}(P^{\prime})+\bar{\sigma}^{\mu}\tilde
T_aS_{34}(P)\bar{\sigma}^{\nu}\tilde T_bS_{34}(P^{\prime})\Big{]}
\label{selfe}
\end{eqnarray}

Substituting Eq.(\ref{25}) to Eq.(\ref{selfe}), we find that
\begin{eqnarray}
&&{\rm
Tr}\Big{[}\sigma^{\mu}T_aS_{21}(P)\sigma^{\nu}T_bS_{21}(P^{\prime})\Big{]}={\rm
Tr}\Big{[}\sigma^{\mu}\frac{1+{\bm\sigma}\cdot\hat{\bf p}}{2}\sigma^{\nu}\frac{1+{\bm\sigma}\cdot\hat{\bf p}^{\prime}}{2}\Big{]}w_1({\bf p},\bf p^{\prime}), \nonumber \\
&&{\rm
Tr}\Big{[}\bar\sigma^{\mu}T_aS_{12}(P)\bar\sigma^{\nu}T_bS_{12}(P^{\prime})\Big{]}={\rm
Tr}\Big{[}\bar\sigma^{\mu}\frac{1-{\bm\sigma}\cdot\hat{\bf p}}{2}\bar\sigma^{\nu}\frac{1-{\bm\sigma}\cdot\hat{\bf p}^{\prime}}{2}\Big{]}w^{\prime}_1({\bf p},{\bf p}^{\prime}), \nonumber \\
&&{\rm Tr}\Big{[}\sigma^{\mu}T_aS_{23}(P)\sigma^{\nu}\tilde
T_bS_{41}(P^{\prime})\Big{]}=\frac{1}{2}{\rm
Tr}\Big{[}\sigma^{\mu}{\bm\sigma}\cdot{\hat{\bf e}}_+(\hat{\bf
p})\phi(\nu,{\bf p})\frac{1-{\bm\sigma}\cdot\hat{\bf
p}}{2}\sigma^{\nu}
{\bm\sigma}\cdot{\hat{\bf e}}_-(\hat{\bf p}^\prime)\phi^*(\nu^\prime,{\bf p}^{\prime})\frac{1+{\bm\sigma}\cdot\hat{\bf p}^{\prime}}{2}\Big{]}w_2({\bf p},{\bf p}^{\prime}), \nonumber\\
&&{\rm Tr}\Big{[}\bar\sigma^{\mu}T_aS_{14}(P)\bar\sigma^{\nu}\tilde
T_bS_{32}(P^{\prime})\Big{]}=\frac{1}{2}{\rm
Tr}\Big{[}\bar\sigma^{\mu}{\bm\sigma}\cdot{\hat{\bf e}}_-(\hat{\bf
p})\phi(\nu,-{\bf p})\frac{1+{\bm\sigma}\cdot\hat{\bf
p}}{2}\bar\sigma^{\nu}
{\bm\sigma}\cdot{\hat{\bf p}}_+(\hat{\bf p}^\prime)\phi^*(\nu^\prime,-{\bf p}^{\prime})\frac{1-{\bm\sigma}\cdot\hat{\bf p}^{\prime}}{2}\Big{]}w^{\prime}_2({\bf p},{\bf p}^{\prime}), \nonumber\\
&&{\rm Tr}\Big{[}\sigma^{\mu}\tilde
T_aS_{41}(P)\sigma^{\nu}T_bS_{23}(P^{\prime})\Big{]}=\frac{1}{2}{\rm
Tr}\Big{[}\sigma^{\mu}{\bm\sigma}\cdot{\hat{\bf e}}_-(\hat{\bf
p})\phi^*(\nu,{\bf p})\frac{1+{\bm\sigma}\cdot\hat{\bf
p}}{2}\sigma^{\nu}
{\bm\sigma}\cdot{\hat{\bf e}}_+(\hat{\bf p}^\prime)\phi(\nu^\prime,{\bf p}^{\prime})\frac{1-{\bm\sigma}\cdot\hat{\bf p}^{\prime}}{2}\Big{]}w_3({\bf p},{\bf p}^{\prime}), \nonumber\\
&&{\rm Tr}\Big{[}\bar\sigma^{\mu}\tilde
T_aS_{32}(P)\bar\sigma^{\nu}T_bS_{14}(P^{\prime})\Big{]}=\frac{1}{2}{\rm
Tr}\Big{[}\bar\sigma^{\mu}{\bm\sigma}\cdot{\hat{\bf e}}_+(\hat{\bf
p})\phi^*(\nu,-{\bf p})\frac{1-{\bm\sigma}\cdot\hat{\bf
p}}{2}\bar\sigma^{\nu}
{\bm\sigma}\cdot{\hat{\bf e}}_-(\hat{\bf p}^\prime)\phi(\nu^\prime,-{\bf p}^{\prime})\frac{1+{\bm\sigma}\cdot\hat{\bf p}^{\prime}}{2}\Big{]}w^{\prime}_3({\bf p},{\bf p}^{\prime}), \nonumber\\
&&{\rm Tr}\Big{[}\sigma^{\mu}\tilde T_aS_{43}(P)\sigma^{\nu}\tilde
T_bS_{43}(P^{\prime})\Big{]}={\rm
Tr}\Big{[}\sigma^{\mu}\frac{1-{\bm\sigma}\cdot\hat{\bf p}}{2}\sigma^{\nu}\frac{1-{\bm\sigma}\cdot\hat{\bf p}^{\prime}}{2}\Big{]}w_4({\bf p},{\bf p}^{\prime}), \nonumber\\
&&{\rm Tr}\Big{[}\bar\sigma^{\mu}\tilde
T_aS_{34}(P)\bar\sigma^{\nu}\tilde T_bS_{34}(P^{\prime})\Big{]}={\rm
Tr}\Big{[}\bar\sigma^{\mu}\frac{1+{\bm\sigma}\cdot\hat{\bf
p}}{2}\bar\sigma^{\nu}\frac{1+{\bm\sigma}\cdot\hat{\bf
p}^{\prime}}{2}\Big{]}w^{\prime}_4({\bf p},{\bf p}^{\prime})
\label{ftrace}
\end{eqnarray}
The trace over color space can be readily performed and we have
\begin{equation}
w_{1,4}({\bf p},{\bf p}^{\prime})=w^{\prime}_{1,4}(-{\bf p},-{\bf
p}^{\prime})=\left\{\begin{array}{ll}
\begin{gathered}
\frac{1}{2}\frac{i\nu\pm\xi}{\nu^2+\xi^2+\Delta^2}\frac{i\nu^\prime\pm\xi^{\prime}}{\nu^{\prime 2}+\xi^{\prime 2}+\Delta^{\prime 2}},\hspace{3.3cm}a=b=1,2,3\\
\end{gathered}
\hfill\\
\begin{gathered}
-\frac{1}{4}\Big{[}\frac{1}{i\nu\mp\xi}\frac{i\nu^{\prime}\pm\xi^{\prime}}{\nu^{\prime
2}+\xi^{\prime 2}+\Delta^{\prime 2}}+
\frac{1}{\nu^{\prime}\mp\xi^{\prime}}\frac{i\nu\pm\xi}{\nu^2+\xi^2+\Delta^2}\Big{]},\hspace{0.2cm}a=b=4,...,7 \\
\end{gathered}
\hfill\\
\begin{gathered}
\frac{1}{6}\frac{i\nu\pm\xi}{\nu^2+\xi^2+\Delta^2}\frac{i\nu^{\prime}\pm\xi^{\prime}}{\nu^{\prime
2}+\xi^{\prime 2}+\Delta^{\prime
2}}+\frac{1}{3}\frac{1}{i\nu\mp\xi}\frac{1}{i\nu^{\prime}\mp\xi^{\prime}},\hspace{0.45cm}a=b=8
\end{gathered}
\end{array}
\right. \label{w1}
\end{equation}
and
\begin{equation}
w_2({\bf p},{\bf p}^{\prime})=w^{\prime}_2(-{\bf p},-{\bf
p}^{\prime})=w_3({\bf p},{\bf p}^{\prime})=w^{\prime}_3(-{\bf
p},-{\bf p}^{\prime})=\left\{\begin{array}{ll}
\begin{gathered}
-\frac{1}{2}\frac{1}{\nu^2+\xi^2+\Delta^2}\frac{1}{\nu^{\prime 2}+\xi^{\prime 2}=\Delta^{\prime 2}},\hspace{0.15cm}a=b=1,2,3\\
\end{gathered}
\hfill\\
\begin{gathered}
0,\hspace{4.95cm}a=b=4,...,7\\
\end{gathered}
\hfill\\
\begin{gathered}
\frac{1}{6}\frac{1}{\nu^2+\xi^2+\Delta^2}\frac{1}{\nu^{\prime
2}+\xi^{\prime 2}+\Delta^{\prime 2}},\hspace{0.5cm}a=b=8
\end{gathered}
\end{array}
\right.\label{w2}
\end{equation}
where we have defined $\xi=p-\mu$, $\Delta^2=|\phi(\nu,{\bf
p})|^2$. The $\pm$ on the RHS of Eq.(\ref{w1}) correspond to $w_1$
and $w_4$ respectively. In weak coupling, the dominant
contribution arise from the quasi-particles, therefore we have
ignored that from the quasi-antiparticles. The trace over Pauli
matrices on the RHS of Eq.(\ref{ftrace}) are evaluated. We have
\begin{eqnarray}
{\rm Tr}[\sigma^0\frac{1\pm{\bm\sigma}\cdot\hat{\bf
p}}{2}\sigma^0\frac{1\pm{\bm\sigma}\cdot\hat{\bf p}^{\prime}}{2}]=
{\rm Tr}[\bar\sigma^0\frac{1\pm{\bm\sigma}\cdot\hat{\bf
p}}{2}\bar\sigma^0\frac{1\pm{\bm\sigma}\cdot\hat{\bf
p}^{\prime}}{2}]=\frac{1}{2}(1+\hat{\bf p}\cdot\hat{\bf
p}^{\prime})\\
{\rm Tr}[\sigma^i\frac{1\pm{\bm\sigma}\cdot\hat{\bf
p}}{2}\sigma^i\frac{1\pm{\bm\sigma}\cdot\hat{\bf p}^{\prime}}{2}]=
{\rm Tr}[\bar\sigma^i\frac{1\pm{\bm\sigma}\cdot\hat{\bf
p}}{2}\bar\sigma^i\frac{1\pm{\bm\sigma}\cdot\hat{\bf
p}^{\prime}}{2}]=\frac{1}{2}(3-\hat{\bf p}\cdot\hat{\bf
p}^{\prime})\\
\hat{k}_i\hat{k}_j{\rm
Tr}[\sigma^i\frac{1\pm{\bm\sigma}\cdot\hat{\bf
p}}{2}\sigma^j\frac{1\pm{\bm\sigma}\cdot\hat{\bf p}^{\prime}}{2}]=
\hat{k}_i\hat{k}_j{\rm
Tr}[\bar\sigma^i\frac{1\pm{\bm\sigma}\cdot\hat{\bf
p}}{2}\bar\sigma^j\frac{1\pm{\bm\sigma}\cdot\hat{\bf
p}^{\prime}}{2}]=\frac{1}{2}[1-\hat{\bf p}\cdot\hat{\bf
p}^{\prime}+2\hat{\bf k}\cdot\hat{\bf p}\hat{\bf k}\cdot\hat{\bf
p}^{\prime}].
\end{eqnarray}
It follows from Eq.(\ref{eigenvalue}) that
\begin{equation}
{\bm\sigma}\cdot{\hat{\bf e}}_\pm(\hat{\bf
p})\frac{1\mp{\bm\sigma}\cdot\hat{\bf p}}{2} =\hat{\bf
e}_\pm(\hat{\bf p}).\label{39}
\end{equation}
and the trace in Eq.(\ref{ftrace}) is simplified. The terms we need
are
\begin{eqnarray}
{\rm Tr}[\sigma^0{\bm\sigma}\cdot{\hat{\bf e}}_+(\hat{\bf p})\phi(\nu,{\bf
p})\frac{1-{\bm\sigma}\cdot\hat{\bf
p}}{2}\sigma^0{\bm\sigma}\cdot{\hat{\bf e}}_-(\hat{\bf p}^\prime)\phi^*(\nu^{\prime},{\bf
p}^{\prime})\frac{1+{\bm\sigma}\cdot\hat{\bf
p}^{\prime}}{2}]=2\phi(\nu,{\bf p})\phi^*(\nu^\prime,{\bf
p}^{\prime})\hat{\bf e}_+(\hat{\bf
p})\cdot\hat{\bf e}_-(\hat{\bf p}^{\prime})\\
{\rm Tr}[\sigma^i{\bm\sigma}\cdot{\hat{\bf e}}_+(\hat{\bf p})\phi(\nu,{\bf
p})\frac{1-{\bm\sigma}\cdot\hat{\bf
p}}{2}\sigma^i{\bm\sigma}\cdot{\hat{\bf e}}_-(\hat{\bf p}^\prime)\phi^*(\nu^{\prime},{\bf
p}^{\prime})\frac{1+{\bm\sigma}\cdot\hat{\bf p}^{\prime}}{2}]=
-2\phi(\nu,{\bf p})\phi^*(\nu^\prime,{\bf
p}^{\prime})\hat{\bf e}_+(\hat {\bf p})\cdot\hat{\bf e}_-(\hat{\bf
p}^{\prime})
\end{eqnarray}
and
\begin{eqnarray}
\nonumber & &\hat{k}_i\hat{k}_j{\rm
Tr}[\sigma^i{\bm\sigma}\cdot{\hat{\bf e}}_+(\hat{\bf p})\phi(\nu,{\bf
p})\frac{1-{\bm\sigma}\cdot\hat{\bf
p}}{2}\sigma^j{\bm\sigma}\cdot{\hat{\bf e}}_-(\hat{\bf p}^\prime)\phi^*(\nu^\prime,{\bf
p}^{\prime})\frac{1+{\bm\sigma}\cdot\hat{\bf p}^{\prime}}{2}]\\
&=&\frac{4(pp^{\prime}-{\bf p}\cdot{\bf p}^{\prime})}{({\bf
p}-{\bf p}^{\prime})^2}\phi(\nu,{\bf
p})\phi^*(\nu^\prime,{\bf p}^{\prime})\hat{\bf e}_+(\hat{\bf
p})\cdot\hat{\bf e}_-(\hat{\bf p}^{\prime}) -2\phi(\nu,{\bf
p})\phi^*(\nu^\prime,{\bf p}^{\prime})\hat{\bf e}_+(\hat{\bf
p})\cdot\hat{\bf e}_-(\hat{\bf p}^{\prime})
\end{eqnarray}
The other cases in Eq.(\ref{ftrace}) can be evaluated along the same
lines and we do not write them down explicitly here.

We now turn to the Hartree-Fock energy density in Eq.(\ref{HF})
and divide it into two parts
\begin{equation}
\Gamma_{HF}=-\frac{1}{2\Omega}{\rm
Tr}[D_n\delta\Pi]=\Gamma_m+\Gamma_e
\end{equation}
where the first term comes from magnetic gluons and the second
term from electric gluons and we have
\begin{subequations}
\begin{eqnarray}
\Gamma_m=-\frac{1}{4}g^2\sum_{P,P^{\prime}}D_t(K)(\delta_{ij}-\hat
k_i\hat k_j){\rm
Tr}\Big{[}\hat{\Gamma}_{a}^{i}S(P)\hat{\Gamma}_{a}^{j}S(P^{\prime})\Big{]}\\
\Gamma_e=-\frac{1}{4}g^2\sum_{P,P^{\prime}}D_l(K){\rm
Tr}\Big{[}\hat{\Gamma}_{a}^{0}S(P)\hat{\Gamma}_{a}^{0}S(P^{\prime})\Big{]}
\end{eqnarray}
\end{subequations}

As we discussed in our previous paper\cite{BDH}, the net
contribution from Eq.(\ref{w1}) and that from the diagonal block
of the quark self-energy matrix is to add a wave-function
renormalization factor in the gap equation\cite{WJH3, WJH,
WJH2,Qun}, which is subleading order. It will not interfere with
the angular dependence of the gap function as we have already seen
in our previous paper\cite{BDH}. Therefore, the contribution from Eq.(\ref{w1}) will be
ignored and only that from Eq.(\ref{w2}) will be considered here. We find
\begin{eqnarray}
\nonumber
\Gamma_m=&-&\frac{2}{3}g^2\sum_{P,P^{\prime}}D_t(K)\frac{pp^{\prime}-{\bf
p}\cdot{\bf p}^{\prime}}{({\bf p}-{\bf
p}^{\prime})^2}\frac{\phi(\nu,{\bf p})\phi^*(\nu^\prime,{\bf
p}^{\prime})\hat{\bf e}_+(\hat{\bf p})\cdot\hat{\bf e}_-(\hat{\bf
p}^{\prime})+c.c.}{(\nu^2+\xi^2+\Delta^2)(\nu^{\prime
2}+\xi^{\prime 2}+\Delta^{\prime 2})}\\
&+&\{{\rm a\hspace{0.15cm} term\hspace{0.15cm} with}
\hspace{0.15cm}{\bf p}\rightarrow-{\bf p},{\bf
p}^{\prime}\rightarrow-{\bf p}^{\prime}\}
\end{eqnarray}
and
\begin{equation}
\Gamma_e=-\frac{1}{3}g^2\sum_{P,P^{\prime}}D_l(K)\frac{\phi(\nu,{\bf
p})\phi^*(\nu^{\prime},{\bf p}^{\prime})\hat{\bf e}_+(\hat{\bf
p})\cdot\hat{\bf e}_-(\hat{\bf
p}^{\prime})+c.c.}{(\nu^2+\xi^2+\Delta^2)(\nu^{\prime 2}+\xi^{\prime
2}+\Delta^{\prime 2})}+\{{\rm a\hspace{0.15cm} term\hspace{0.15cm}
with} \hspace{0.15cm}{\bf p}\rightarrow-{\bf p},{\bf
p}^{\prime}\rightarrow-{\bf p}^{\prime}\}
\end{equation}
In weak coupling, the approximation $p\simeq p^{\prime}\simeq\mu$
is justified, and we can perform the integral over $\xi$ by
picking up the the poles in the diquark propagator. Assuming
further that the gap function depends only on the energy and the
orientation of the momentum, we find
\begin{equation}
\Gamma_{HF}=-\frac{1}{6}g^2\mu^4\int\frac{d\nu}{2\pi}\int\frac{d\nu^{\prime}}{2\pi}\int\frac{d^2\hat{\bf
p}}{(2\pi)^2}\int\frac{d^2\hat{\bf
p}^{\prime}}{(2\pi)^2}\Big{[}D_t(K)+D_l(K)\Big{]}
\frac{\phi(\nu,{\bf p})\phi^*(\nu^\prime,{\bf
p}^{\prime})\hat{\bf e}_+(\hat{\bf p})\cdot\hat{\bf e}_-(\hat{\bf
p}^{\prime})+c.c.}{\sqrt{\nu^2+\Delta^2}\sqrt{\nu^{\prime
2}+\Delta^{\prime 2}}}
\end{equation}

In order to evaluate the 3rd and 4th terms on the RHS of Eq.(\ref{28}),
we employ the identities ${\rm Tr ln}S^{-1}={\rm ln Det}S^{-1}$
and
\begin{equation}
{\rm Det}\left(\begin{array}{cc} A & B\\
C & D
\end{array}
\right)={\rm Det}(-CB+CAC^{-1}D)={\rm Det}(-BC+BDB^{-1}A)
\end{equation}
with $A$, $B$, $C$ and $D$ arbitrary matrices and $B,C$
invertible. We have
\begin{equation}
\frac{1}{2\Omega}({\rm Tr ln}S^{-1}-{\rm Tr
ln}S_0^{-1})=2\mu^2\int\frac{d\nu}{2\pi}\int\frac{d^2\hat{ \bf
p}}{(2\pi)^2}(|\nu|-\sqrt{\nu^2+\Delta^2})
\end{equation}
The evaluation of the last term on the RHS of Eq.(\ref{28}) can be
readily performed by using the full quark propagator and we end up
with
\begin{equation}
\frac{1}{2\Omega}{\rm
Tr}(S_0^{-1}S-1)=2\mu^2\int\frac{d\nu}{2\pi}\int\frac{d^2\hat{\bf
p}}{(2\pi)^2}\frac{\Delta^2}{\sqrt{\nu^2+\Delta^2}}
\end{equation}
Therefore, the condensation energy density reads
\begin{eqnarray}
\nonumber F=&-&\frac{3{\bar g}^2\mu^4}{64\pi^4}\int d\nu\int
d\nu^{\prime}\int d^2\hat{\bf p}\int d^2\hat{\bf
p}^{\prime}V(\nu-\nu^{\prime},\hat{\bf p}\cdot\hat{\bf
p}^{\prime})\frac{\phi(\nu,{\bf p})\phi^*(\nu^\prime,{\bf
p}^{\prime})\hat{\bf e}_+(\hat{\bf p})\cdot\hat{\bf e}_-(\hat{\bf
p}^{\prime})+c.c.}{\sqrt{\nu^2+\Delta^2}\sqrt{\nu^{\prime
2}+\Delta^{\prime 2}}}\\
&+&\frac{\mu^2}{4\pi^3}\int d\nu\int d^2\hat{\bf
p}\Big{(}|\nu|-\frac{\nu^2}{\sqrt{\nu^2+\Delta^2}}\Big{)}
\label{56}
\end{eqnarray}
where, the pairing potential $V$ contains the contributions from
both magnetic and electric gluons and reads
\begin{equation}
V(\nu-\nu^{\prime},\hat{\bf p}\cdot\hat{\bf
p}^{\prime})=D_t(\nu-\nu^{\prime},{\bf p}-{\bf
p}^{\prime})+D_l(\nu-\nu^{\prime},{\bf p}-\hat{\bf
p}^{\prime})
\end{equation}
with ${\bf p}-{\bf p}^{\prime}=\mu(\hat{{\bf p}}-\hat{{\bf
p}}^{\prime})$. In the energy-momentum region for Cooper pairing,
$|\nu-\nu^\prime|<<|{\bf p}-{\bf p}^\prime|$, $D_t$ is dominated
by Landau damping and $D_l$ by Debye screening. We have
\begin{equation}
V(\nu-\nu^{\prime},\hat{\bf p}\cdot\hat{\bf p}^{\prime})
=\frac{1}{({\bf p}-{\bf
p}^\prime)^2+\frac{\pi}{4}m_D^2\frac{|\nu-\nu^\prime|}{|{\bf
p}-{\bf p}^{\prime}|}} +\frac{1}{({\bf p}-{\bf p}^\prime)^2+m_D^2}
\label{HDL}
\end{equation}
where the Debye mass square $m_D^2=\frac{N_fg^2\mu^2}{2\pi^2}$.

To obtain the gap equation, we introduce the quantity,
\begin{equation}
\zeta=\frac{\phi}{\sqrt{\nu^2+\Delta^2}}
\end{equation}
The minimization of the condensation energy density with respect
to $\zeta$ gives rise to the gap equation
\begin{equation}
\phi(\nu,\hat{\bf p})=\frac{3{\bar g}^2\mu^2}{4\pi}\int
d\nu^{\prime}\int d^2\hat{\bf p}^{\prime}V(\nu-\nu^{\prime},
\hat{\bf p}\cdot\hat{\bf
p}^{\prime})\frac{\phi(\nu^{\prime},\hat{\bf
p}^{\prime})\hat{\bf e}_+(\hat{\bf p}^{\prime})\cdot\hat{\bf
e}_-(\hat{\bf p})}{\sqrt{\nu^{\prime 2}+\Delta^{\prime 2}}}.
\label{gequation}
\end{equation}
The forward singularity corresponds to the divergence of the angular integration when
the coupling $g$ in the denominators of (\ref{HDL}) is set to zero. To isolate out
the singularity, we write the integral on RHS of (\ref{gequation}) as
\begin{eqnarray}
\nonumber &&\int d\nu^{\prime}\int d^2\hat{\bf
p}^{\prime}V(\nu-\nu^{\prime}, \hat{\bf p}\cdot\hat{\bf
p}^{\prime})\frac{\phi(\nu^{\prime},\hat{\bf
p})}{\sqrt{\nu^{\prime 2}+|\phi(\nu^\prime, \hat{{\bf p}})|^2}}\\
&+&\int d\nu^{\prime}\int d^2\hat{\bf
p}^{\prime}V(\nu-\nu^{\prime}, \hat{\bf p}\cdot\hat{\bf
p}^{\prime}) \Big[\frac{\phi(\nu^{\prime},\hat{\bf
p}^{\prime})\hat{\bf e}_+(\hat{\bf p}^{\prime})\cdot\hat{\bf
e}_-(\hat{\bf p})}{\sqrt{\nu^{\prime 2}+|\phi(\nu^\prime,
\hat{{\bf p}}^\prime)|^2}} -\frac{\phi(\nu^{\prime},\hat{\bf p})}
{\sqrt{\nu^{\prime 2}+|\phi(\nu^\prime, \hat{{\bf p}})|^2}}\Big].
\label{decomp}
\end{eqnarray}
The coupling $g$ in the second integral can be set to zero without introducing
divergence. So we did and the gap equation reduces to
\begin{eqnarray}
\nonumber \phi(\nu,\hat{\bf p})=&&\int_{-\omega_0}^{\omega_0}
d\nu^{\prime}\Big{\{}\frac{1}{2}{\rm
ln}\frac{\omega_c}{|\nu-\nu^{\prime}|}
\frac{\phi(\nu^{\prime},\hat{\bf p})}{\sqrt{\nu^{\prime 2}+|\phi(\nu^{\prime},\hat{\bf p})|^2}}\\
&&+\frac{3\bar g^2}{4\pi}\int d^2\hat{\bf
p}^{\prime}\frac{1}{1-\hat{\bf p}\cdot\hat{\bf
p}^{\prime}}\Big{[}\frac{\phi(\nu^{\prime},\hat{\bf
p}^{\prime})\hat{\bf e}_+(\hat{\bf p}^{\prime})\cdot\hat{\bf
e}_-(\hat{\bf p})}{\sqrt{\nu^{\prime 2}+|\phi(\nu^\prime,\hat{\bf
p}^\prime)|^2 }}-\frac{\phi(\nu^{\prime},\hat{\bf p})}
{\sqrt{\nu^{\prime 2}+|\phi(\nu^\prime,\hat{\bf
p})|^2}}\Big{]}\Big{\}},
\label{redgapeq}
\end{eqnarray}
where we have carried out the $\hat{\bf p}^\prime$ integral of the
first term of (\ref{decomp}) and have inintroduced an infrared
cutoff $\omega_0\sim g\mu$.

\section{The integral equation for angular dependence and its solution}

In the expansion Eq.(\ref{circular}), the circular polarization
basis $\hat{\bf e}_\pm$ carry singularity either at the north pole
or at the south pole of the spherical coordinates. The coefficient
functions that remove this singularity are the Wigner D-functions,
instead of the ordinary spherical harmonics. The Winger D-function
corresponds to the irreducible representation of the rotation
group\cite{ZJY} and is defined by
\begin{equation}
D_{mm'}^J(\alpha,\beta,\gamma)={\rm
exp}(-im\alpha)d_{mm'}^J(\beta){\rm
exp}(-im'\gamma)
\end{equation}
with
\begin{equation}
d_{mm'}^J(\beta)=<Jm|e^{-iJ_y\beta}|Jm'>.
\end{equation}
Its complex conjugate, $D_{mm'}^{J*}(\alpha,\beta,\gamma)$, is the
wave function of a symmetric top of definite $J^2$, $J_z$ and
${\bf J}\cdot\hat\zeta$ with $\hat\zeta$ the symmetry axis. A
diquark pair of opposite helicity of a definite angular momentum
represents a parallel case with $\hat\zeta$ of the symmetric top
replaced by direction of the relative momentum of the pair,
$\hat{\textbf{p}}$. Since $\textbf{J}\cdot\hat{\textbf{p}}=\pm 1$,
the wave function of a transverse diquark pair ought to be
proportional to $D_{m\pm 1}^{J*}$. Here, however, there is a
question when carrying this wave function to the diquark pair. A
general rotation is defined by three Euler angles $\alpha, \beta,
\gamma$ but the gap function depends on only $\hat{\bf p}$, which
is defined by the two polar angles $\varphi$ and $\theta$. In what
follows, we shall identify the first two Euler angles with the
polar angles, i.e. $\alpha=\varphi$, $\beta=\theta$, and set
$\gamma=0$. As is shown in the appendix B, the real and imaginary
parts of
$D_{m\pm1}^{J*}(\varphi,\theta,0)\hat{\textbf{e}}_{\pm}(\hat{\textbf{p}})$
are proportional to the vector spherical harmonics, which are free
from the singularities at $\theta=0$ and $\theta=\pi$. Therefore
the partial wave basis of the gap function (\ref{gapfunction}) is
$D_{m1}^{J*}(\varphi,\theta,0)$.

The partial wave amplitude of the pairing potential, $V(\nu-\nu^{\prime},
\hat{\bf p}\cdot\hat{\bf p}^{\prime})\hat{\bf e}_+(\hat{\bf p}^{\prime})\cdot\hat{\bf
e}_-(\hat{\bf p})$, is given by the integral
\begin{eqnarray}
\nonumber &&\int d^2\hat{\bf p}\int d^2{\hat{\bf p}}^\prime
D_{m_11}^J(\varphi,\theta,0) V(\nu-\nu^{\prime},\hat{\bf
p}\cdot\hat{\bf p}^{\prime}) \hat{\bf e}_+(\hat{\bf
p}^{\prime})\cdot\hat{\bf e}_-(\hat{\bf p})
D_{m_21}^{J*}(\varphi,\theta,0)\\
\nonumber &=&\frac{\delta_{m_1m_2}}{2J+1}\int d^2\hat{\bf p}\int
d^2{\hat{\bf p}}^\prime V(\nu-\nu^{\prime},\hat{\bf
p}\cdot\hat{\bf p}^{\prime})
\sum_mD_{m1}^J(\varphi,\theta,0)D_{m1}^{J*}(\varphi,\theta,0)\\
\nonumber&=&\frac{\delta_{m_1m_2}}{2J+1} \int d^2\hat{\bf p}\int
d^2{\hat{\bf p}}^\prime V(\nu-\nu^{\prime},\hat{\bf
p}\cdot\hat{\bf p}^{\prime})
d_{11}^J(\Theta)\cos^2\frac{\Theta}{2}\\
&=&\frac{8\pi^2}{2J+1}\delta_{m_1m_2} \int_{-1}^1d\cos\Theta
V(\nu-\nu^\prime,\cos\Theta)d_{11}^J(\Theta)\cos^2\frac{\Theta}{2}
\label{wignereckart}
\end{eqnarray}
with $\cos\Theta\equiv\hat{\bf p}\cdot\hat{\bf p}^{\prime}$, which corresponds to
expansion coefficients of $V(\nu-\nu^\prime,\cos\Theta)\cos^2\frac{\Theta}{2}$ in
the series of $d_{11}^J(\Theta)$'s. The first equality of (\ref{wignereckart})
follows from the Wigner-Eckart theorem and the second one from the
addition formula (B13)(proved in Appendix B)
\begin{equation}
\sum_mD_{m,1}^{J^*}(\varphi,\theta,0)D_{m,1}^J(\varphi^{\prime},\theta^{\prime},0)
\hat{\bf e}_+(\hat{\bf p})\cdot\hat{\bf e}_-(\hat{\bf p}^{\prime})
=d_{11}^J(\Theta)\cos^2\frac{\Theta}{2}.
\end{equation}
We have\cite{WJH2, WJH,WJH3}
\begin{equation}
V(\nu-\nu^{\prime}, \hat{\bf p}-\hat{\bf
p}^{\prime})\cos^2\frac{\Theta}{2}=\frac{1}{6\mu^2}{\rm
ln}\frac{\omega_c}{|\nu-\nu^{\prime}|}\sum_{J=1}^\infty(2J+1)d_{11}^J(\Theta)
+\frac{1}{2\mu^2}\sum_{J=1}^\infty(2J+1)c_J^\prime d_{11}^J(\Theta)
\label{gapeq}
\end{equation}
where, $\omega_c=\frac{1024\sqrt{2}\pi^4\mu}{g^5}$ and
$c_J^\prime$ is given by
\begin{equation}
c_J^\prime=\int^{1}_{-1}d\cos\Theta\frac{d_{11}^J(\Theta)\cos^2\frac{\Theta}{2}-1}{1-\cos\Theta}
=\left\{\begin{array}{ll}
\begin{gathered}
-\frac{3}{2},\hspace{2.88cm}\hbox{for $J=1$}\\
\end{gathered}
\hfill\\
\begin{gathered}
-\sum_n^J\frac{1}{n}+\frac{1}{2J(J+1)},\hspace{0.2cm}\hbox{for $J>1$}\\
\end{gathered}
\end{array}
\right.
\end{equation}
Unlike the magnitude of the angular momentum, $J$, that is mixed
in the solution because of the nonlinearity of the gap equation. The
azimuthal quantum number, $J_z=m$, remains conserved. For a gap function of a
definite $m$,
\begin{equation}
\phi(\nu,\hat{\bf p})=\psi_m(\nu,\cos\theta)e^{im\phi}
\end{equation}
with $\psi_m(\nu,\cos\theta)=$ real, the gap equation (\ref{redgapeq}) implies the
following integral equation satisfied by $\psi_m(\nu,\cos\theta)$
\begin{eqnarray}
\nonumber \psi_m(\nu,x)=&&\frac{1}{2}\bar g^2\int_{0}^{\omega_0}
d\nu^{\prime}\Big{\{}\frac{1}{2}({\rm
ln}\frac{\omega_c}{|\nu-\nu^{\prime}|}+{\rm
ln}\frac{\omega_c}{\nu+\nu^{\prime}})
\frac{\psi_m(\nu^{\prime},x)}{\sqrt{\nu^{\prime
2}+\psi_m^2(\nu^{\prime},x)}}\\
&&+3\int_{-1}^{1}dx^\prime\frac{1}{|x-x^\prime|}
\Big{[}\frac{{\cal
I}_m(x,x^\prime)\psi_m(\nu^{\prime},x^{\prime})}
{\sqrt{\nu^{\prime
2}+\psi_m^2(\nu^{\prime},x^{\prime})}}-\frac{\psi(\nu^{\prime},x)}
{\sqrt{\nu^{\prime 2}+\psi_m^2(\nu^{\prime},x)}}\Big{]}\Big{\}},
\label{gapequation}
\end{eqnarray}
where $x=\cos\theta$, $x^\prime=\cos\theta^\prime$ and
${\cal I}_m(x,x^\prime)\equiv |x-x^\prime|I_m(x,x^\prime)$ with
\begin{equation}
I_m(x,x^{\prime})=\frac{1}{2\pi}\int_{0}^{2\pi}d\varphi\frac{\hat{\bf
e}_+(\hat{ \bf p}^{\prime})\cdot\hat{\bf e}_-(\hat{\bf
p})e^{im(\varphi^{\prime}-\varphi)}}{1-\hat{\bf p}\cdot\hat{\bf
p}^{\prime}}.
\end{equation}
Substituting the explicit expression of $\hat{\bf e}_+(\hat{\bf
p}^{\prime})\cdot\hat{\bf e}_-(\hat{\bf p})$,  we find that
$I_m(x,x^\prime)$ is real and
\begin{equation}
I_{-m}(x,x^\prime)=I_m(-x,-x^\prime).
\label{symmetry}
\end{equation}

The reduction of the gap equation (\ref{gapequation}) is parallel to that for the
longitudinal pairing case. We shall outline only the main steps here and refer the interested
reader to the details in the section IV of \cite{BDH}. We start with the approximation
in \cite{D}
\begin{equation}
{\rm ln}\frac{\omega_c}{|\nu-\nu^{\prime}|}\simeq{\rm
ln}\frac{\omega_c}{|\nu_{>}|}
\end{equation}
with $\nu_{>}=\rm max(\nu,\nu^{\prime})$ and introducing the new variable
\begin{equation}
\xi={\rm ln}\frac{\omega_c}{\nu}.
\end{equation}
Twice differentiation of Eq.(\ref{gapequation}) with respect to
$\xi$ yields an ordinary differential equation
\begin{equation}
\frac{d^2\psi_m(\xi,x)}{d\xi^2}+\frac{{\bar
g}^2\psi_m(\xi,x)}{\sqrt{1+\frac{\psi_m^2(\xi,x)}{\omega_c^2}e^{2\xi}}}=0\label{gapeq4}
\end{equation}
together with the boundary condition
\begin{equation}
\frac{d\psi_m(\xi,x)}{d\xi}=0\label{bcondition}
\end{equation}
as $\xi\rightarrow\infty$, valid for all $x$. The solution of
Eq.(\ref{gapeq4}) up to the subleading order reads:
\begin{equation}
\psi_m(\xi,x)=\Delta_0f_m(x)\Big[A(\xi,x)u(\xi,x)+B(\xi,x)v(\xi,x)\Big]\label{phiax}
\end{equation}
where
\begin{equation}
u(\xi,x)=\left\{\begin{array}{ll}
\begin{gathered}\cos\bar g[b(x)-\xi],\hspace{0.2cm}\hbox{for $\xi<b(x)$}\\
\end{gathered}
\hfill\\
\begin{gathered}1,\hspace{2.1536cm}\hbox{for $\xi>b(x)$}\\
\end{gathered}
\end{array}
\right.
\end{equation}
\begin{equation}
v(\xi,x)=\left\{\begin{array}{ll}
\begin{gathered}-\sin\bar g[b(x)-\xi],\hspace{0.2cm}\hbox{for $\xi<b(x)$}\\
\end{gathered}
\hfill\\
\begin{gathered}\bar g\xi,\hspace{2.255cm}\hbox{for $\xi>b(x)$}\\
\end{gathered}
\end{array}
\right.
\end{equation}
and
\begin{subequations}
\begin{equation}
A(\xi,x)=1+\bar{g}\int\limits_{\xi}^{\infty}d\xi^{\prime}\Big[\theta(b-\xi^{\prime})-\frac{1}
{\sqrt{1+u^2(\xi^{\prime},x)e^{2[\xi^{\prime}-b(x)]}}}\Big]v(\xi^{\prime},x)u(\xi^\prime,x)
\label{50a}
\end{equation}
\begin{equation}
B(\xi,x)=\bar{g}\int\limits_{\xi}^{\infty}d\xi^{\prime}\Big[\theta(b-\xi^{\prime})-\frac{1}
{\sqrt{1+u^2(\xi^{\prime},x)e^{2[\xi^{\prime}-b(x)]}}}\Big]u^2(\xi^{\prime},x).
\label{50b}
\end{equation}
\end{subequations}
Here, we have introduced the energy gap of 2SC
\begin{equation}
\Delta_0=\frac{2048\sqrt{2}\pi^4\mu}{N_f^{5/2}g^5}
e^{-\frac{3\pi^2}{\sqrt{2}g}-\frac{\pi^2+4}{8}}
\end{equation}
and the quantity $b(x)=\ln\frac{\omega_c}{\Delta_0|f_m(x)|}$. The
angle dependent function $f_m(x)$ will be determined below. We notice that
at zero Matsubara energy, $\xi\to\infty$,
$\psi_m(0,x)=\Delta_0f_m(x)$. Substituting Eq.(\ref{phiax}) back
into Eq.(\ref{gapequation}), we derive the integral equation for
$f_m(x)$
\begin{equation}
f_m(x){\rm
ln}|f_m(x)|+3\int_{-1}^{1}dx^{\prime}\frac{1}{|x-x^{\prime}|}\Big{[}f_m(x)-{\cal I}_m(x,x')f_m(x^{\prime})\Big{]}=0.
\label{73}
\end{equation}
The solution of (\ref{73}) contains all partial waves with $J\ge{\rm max}(1,|m|)$
and we shall concentrate on most favored pairing channels,
$m=0,\pm1$. In ${}^3{\rm He}$, $m=0$ refers to the
polar phase and $m=\pm1$ to the A phase. We have
\begin{equation}
I_0(x,x^\prime)=-\frac{1+xx^{\prime}}{2\sqrt{1-x^2}\sqrt{1-x^{\prime
2}}}+\frac{1}{2|x-x^{\prime}|}\Big{(}\frac{\sqrt{1-x^{\prime
2}}}{\sqrt{1-x^2}}+\frac{\sqrt{1-x^2}}{\sqrt{1-x^{\prime
2}}}\Big{)}
\end{equation}
and
\begin{equation}
I_{\pm1}(x,x^\prime)=\frac{1}{2}\Big{[}\frac{1}{|x-x^{\prime}|}\Big{(}
\frac{1\pm x^{\prime}}{1\pm x}+\frac{1\pm x}{1\pm x^{\prime}}\Big{)}-\frac{1}{1\pm x}-\frac{1}{1\pm x^{\prime}}\Big{]}.
\end{equation}

Because of the symmetry (\ref{symmetry}), given a solution of the integral equation of the azimuthal
number $m$, $f_m(x)$, the function $f_m(-x)$ solves the equation of the azimuthal number
$-m$. For the solution starting with the $J=1$ harmonics, we expect that
\begin{equation}
f_{-m}(x)=f_m(-x)
\end{equation}
so $f_0(x)$ will be an even function of $x$. Since $d_{01}^J(\theta)$ is proportional to the
associate Legendre function $P_J^1(x)$, only the partial waves of odd $J$ contribute to
$f_0(x)$ and the higher harmonics starts from $J=3$. As to $f_{\pm1}(x)$, all higher partial waves with
$J\ge 2$ will be mixed in, so the free energy reduction is expected to be larger than that of $f_0(x)$.

As we have done in \cite{BDH}, this type of integral equation can
be solved by a variational method. Upon the leading order solution
of the gap function and the definition of angular dependent
factor, the condensation energy density Eq.(\ref{56}) turns out to
be a functional of $f(x)$(see appendix B in \cite{BDH}). we have
\begin{equation}
F=\frac{\mu^2\Delta_0^2}{2\pi^2}\mathcal{F}[f_m]
\end{equation}
with
\begin{equation}
\mathcal{F}[f_m(x)]=\int_{-1}^{1}dxf_m^2(x)[{\rm
ln}|f_m(x)|-\frac{1}{2}]+\frac{3}{2}\int_{-1}^{1}dx\int_{-1}^{1}dx^{\prime}\frac{1}{|x-x^{\prime}|}\Big{\{}
f_m^2(x)-2f_m(x)f_m(x^{\prime}){\cal I}_m(x,x')+f_m^2(x^{\prime})\Big{\}}
\label{75}
\end{equation}
It can be easily verified that the variational minimization of
Eq.(\ref{75}) does solve Eq.(\ref{73}). Before the numerical
calculations, we consider a trial function for each case,
$m=0,\pm1$. For $m=0$, the trial function is
\begin{equation}
f_0(x)=c_0d_{01}^1(\theta)=c_0\sqrt{\frac{1-x^2}{2}} \label{76}
\end{equation}
and substitute it into the condensation energy density
Eq.(\ref{75}). The minimization yields
\begin{equation}
c_0=\frac{1}{\sqrt{2}}e^{\frac{5}{6}}e^{-\frac{9}{2}}
\end{equation}
at which
\begin{equation}
\mathcal{F}=-1.214\times 10^{-4}\label{78}
\end{equation}
For $m=\pm1$, the trial function is
\begin{equation}
f_{\pm1}(x)=c_{\pm}d_{\pm11}^1(\cos\theta)=c_{\pm}\frac{(1\pm x)}{2}\label{79}
\end{equation}
The same procedure will give rise to
\begin{equation}
c_\pm=e^{\frac{1}{3}}e^{-\frac{9}{2}}
\end{equation}
at which
\begin{equation}
\mathcal{F}=-8.084\times 10^{-5}\label{81}
\end{equation}
The target functional corresponding to $m=\pm1$ has the same
value, which is also true when the situation goes to angular
momentum mixing. The condensations with $m=\pm1$, therefore, have
the equal priority when the temperature cools down to the critical
temperature.

The trial function Eq.(\ref{76}) and Eq.(\ref{79}) correspond to
what people carried over from the polar phase and A phase of
${}^3{\rm He}$ respectively\cite{T,A}, which contain p-wave only.
The corresponding condensation energy are not optimal as we shall
see in the following, which will be lowered further by including
higher partial waves.

\begin{figure}
\includegraphics{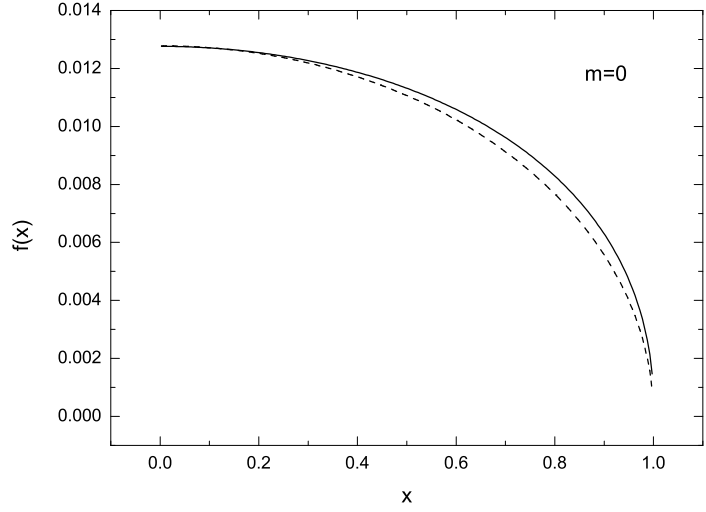}
\caption{\label{fig:eps1} The angular dependence of the gap
function for m=0. The dashed and the solid lines correspond to
trial function Eq.(\ref{76}) and the numerical solution to
Eq.(\ref{73}) respectively.}
\end{figure}

Following the same procedure in section five in \cite{BDH}, we can
obtain the numerical solutions of Eq.(\ref{73}) and the
corresponding condensation energy density. In Fig.\ref{fig:eps1},
we show the angular dependence of the gap function for the case
$m=0$, the dashed line and the solid one are the trial function
and the numerical solution to Eq.(\ref{73}) respectively. They are
depart from each other slightly, however, the condensation energy
density corresponding to each case has qualitative difference. We
find in this case the minimum of target functional
\begin{equation}
\mathcal{F}=-1.236\times 10^{-4}
\end{equation}
which drop from Eq.(\ref{78}) by $1.81$ percent.

\begin{figure}
\includegraphics{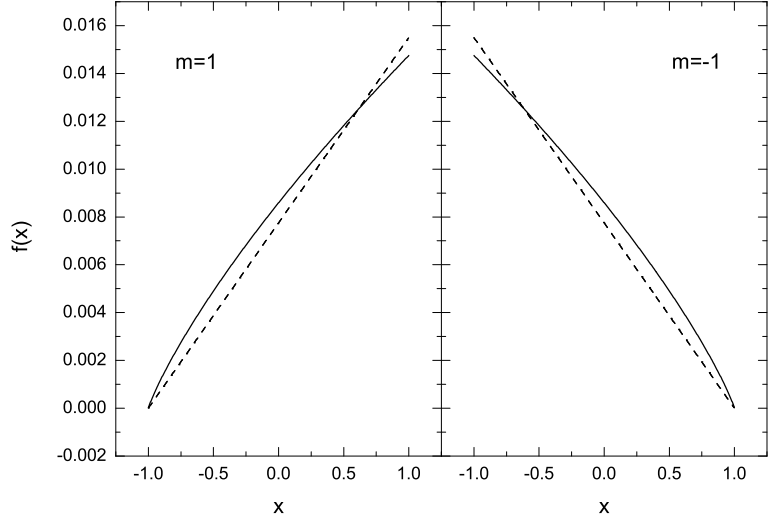}
\caption{\label{fig:eps2} The angular dependence of the gap
function for m=$\pm$1. The dashed and the solid lines correspond
to trial function Eq.(\ref{79}) and the numerical solution to
Eq.(\ref{73}) respectively.}
\end{figure}

The same is true in the case $m=\pm1$. In this case, the trial
functions (dashed lines) and the numerical solutions to
Eq.(\ref{73}) (solid lines) have been plotted in
Fig.\ref{fig:eps2}. We find the minimum of the target functional
corresponding to solid lines
\begin{equation}
\mathcal{F}=-8.494\times 10^{-5}
\end{equation}
which drop from Eq.(\ref{81}) by $5.07$ percent.

As we have found in \cite{BDH}, the drop of the condensation
energy with longitudinal pairing by angular momentum mixing is a
small amount. Here, with transverse pairing, the situation is also
true. With these small amount drop in condensation energy, the
non-spherical pairing of the polar and A-phases in single flavor
CSC remains not energetically more favored than
the spherical pairing state CSL.

It is instructive to examine the angular momentum content of our
solutions according to
\begin{equation}
f_m(\cos\theta)=\sum_{J=1}^{\infty}(2J+1)b_Jd_{m1}^J(\theta)
\end{equation}
for $m=0,\pm1$, we found
\begin{equation}
f_0(\cos\theta)=\frac{1}{\sqrt{2}}e^{5/6}e^{-9/2}[1.0085d_{01}^1(\theta)+0.0426d_{01}^3(\theta)+...]
\end{equation}
and
\begin{equation}
f_{\pm1}(\cos\theta)=e^{1/3}e^{-9/2}[1.0269d_{\pm1
1}^1(\theta)\mp0.0943d_{\pm11}^2(\theta)+...]
\end{equation}

\section{Summary and discussions}
If the pairing potential extends to more than one partial waves, a
non-spherical energy gap cannot be restricted within a single
angular momentum channel because of the nonlinearity of the gap
equation. The corresponding free energy will be brought down by
the mixing. In case of the color superconductivity mediated by
one-gluon exchange, the pairing potential mediated by one-gluon
exchange for all partial waves equal in leading order in QCD
running coupling constant. The angular momentum mixing will occur
to the leading order of the gap function and free energy of all
non-spherical pairing states in the literature needs to be
recalculated. So we examined the angular momentum mixing in this
paper for transverse pairing and in a previous paper for
longitudinal pairing. We derived the nonlinear integral equation
for the angular dependence of the gap function and solved it
numerically. We found that the effect is numerically small, which
may be attributed to the descending pairing strength of higher
partial waves beyond the leading order. Our results on the
condensate energy of various non-spherical pairing states of
single quark flavor are listed in the table I below, in which the
ratios without mixing are also listed for comparison.

While the free energy of all non-spherical pairing states are
lowered by the angular momentum mixing, the amount of the drop is
too small to beat the spherical CSL state of transverse pairing.
The energy balance between the CSL and non-spherical states may be
offset by anisotropy or the violation of time reversal invariance.
In a compact star, the magnetic field and the stellar revolution
provides such a circumstance. The magnetic field, however, is
unlikely to discriminate the non-spherical pairing states from the
spherical ones in the bulk because of the Meissner effect in the
CSC phase \cite{SWR}. In addition, it suppresses all single flavor
pairing because of the critical field. As to the stellar
revolution, the optimal partition of the angular momentum into the
motion of the center of mass and the spin of Cooper pairs remains
to be determined. But the non-spherical condensate will response
differently from the spherical ones because of the preferred
direction in the former.

The slightly complicated non-spherical CSC phase, the transverse
planar phase exhibits high potential to be able to compete CSL
phase since the gain of the condensation energy from the former to
the latter is only 2 percent\cite{A}, falling within the range of
the percentage increment of the condensation energy because of the
angular momentum mixing discovered so far. Technically, the color
structure of the order parameter involves two antisymmetric
Gell-Mann matrices and the present formulation needs to be
generalized. This is currently investigated and the result will be
reported shortly.

\begin{table*}
\caption{\label{tab:table1}The condensation energy of various
non-spherical pairing states in unit of the transverse CSL
condensation energy with or without angular momentum mixing.}
\begin{ruledtabular}
\begin{tabular}{ccccc}
 &\multicolumn{2}{c}{Without mixing}&\multicolumn{2}{c}{With mixing}\\
$F/F_{CSL}$&Longitudinal&Transverse&Longitudinal &Transverse\\
\hline
 Polar & $0.65e^{-12}/e^{-9}$ & 0.88& $0.67e^{-12}/e^{-9}$ & 0.9 \\
A & $0.88e^{-12}/e^{-9}$ & 0.65 & $0.9e^{-12}/e^{-9}$ & 0.68\\
\end{tabular}
\end{ruledtabular}
\end{table*}



\begin{acknowledgments}
We would like to extend our gratitude to  D. Rischke, T.
Sch$\ddot{a}$fer, A. Schmitt for  stimulating discussions. We are
also benefitted from conversations with  M. Huang,  J.R. Li, Q. wang
and P.F. Zhuang. The work of D. F. H. and H. C. R. is supported in
part by NSFC under grant Nos. 10575043, 10735040. The work of D. F.
H. is also supported in part by Educational Committee under under
grants NCET-05-0675 project No. IRT0624.

\end{acknowledgments}

\appendix

\section{The entries of the full quark propagator}
In order to obtain the full quark propagator, we introduce
\begin{equation}
\Gamma_0=\left(\begin{array}{cc} \gamma_0 & 0\\
0 & \gamma_0
\end{array} \right )
\hspace{0.5cm}\rm{and}\hspace{0.5cm} V=\left(\begin{array}{cccc} 1 & 0 & 0 & 0\\
0 & 0 & 1 & 0\\ 0 & 1 & 0 & 0\\ 0 & 0 & 0 & 1
\end{array} \right)
\end{equation}
Using these two matrices, we can transform the inverse quark
propagator into a block diagonal form
\begin{equation}
G(P)=VS^{-1}(P)\Gamma_0V^{-1}=\left( \begin{array}{cc}
T^{-1}(P,\lambda_2) & 0\\
0 & T^{-1}(-P,-\lambda_2)
\end{array}\right)
\end{equation}
where the matrix $T$ is given by
\begin{equation}
T(P,\lambda_2)=\left( \begin{array}{cc}
i\nu+\mu-{\bm\sigma}\cdot\textbf{p} &
\lambda_2{\bm\sigma}\cdot{\bm\Psi}(P)\\
\lambda_2{\bm\sigma}\cdot{\bm\Psi}^*(\bar P) &
i\nu-\mu-{\bm\sigma}\cdot\textbf{p}
\end{array}\right)
\end{equation}
and can be inverted readily by using the projective operators
$P_{\pm}=(1\pm{\bm\sigma}\cdot\hat{\textbf{p}})/2$. Then we have
the full quark propagator
\begin{equation}
S(P)=\Gamma_0V^{-1}G^{-1}(P)V=\left( \begin{array}{cccc} S_{11} &
S_{12} & S_{13} & S_{14}\\ S_{21} & S_{22} & S_{23} & S_{24}\\
S_{31} & S_{32} & S_{33} & S_{34}\\ S_{41} & S_{42} & S_{43} &
S_{44}
\end{array}\right)
\end{equation}
with each entry given by
\begin{subequations}
\begin{eqnarray}
S_{11}=S_{13}=S_{22}=S_{24}=S_{31}=S_{33}=S_{42}=S_{44}=0\\
S_{12}=-\sum_{e\pm}\frac{i\nu-e(p+e\mu)}{\nu^2+(p+e\mu)^2+\lambda^2_2{\bm\sigma}\cdot{\bm\Psi}(-P){\bm\sigma}\cdot{\bm\Psi}^*(-\bar
P)}\frac{1+e{\bm\sigma}\cdot\hat{\textbf{p}}}{2}\\
S_{14}=-\lambda_2{\bm\sigma}\cdot{\bm\Psi}(-P)\sum_{e\pm}\frac{1}{\nu^2+(p-e\mu)^2+\lambda^2_2{\bm\sigma}\cdot{\bm\Psi}^*(-\bar
P){\bm\sigma}\cdot{\bm\Psi}(-
P)}\frac{1+e{\bm\sigma}\cdot\hat{\textbf{p}}}{2}\\
\label{A5d}
S_{21}=-\sum_{e\pm}\frac{i\nu+e(p-e\mu)}{\nu^2+(p-e\mu)^2+\lambda^2_2{\bm\sigma}\cdot{\bm\Psi}(P){\bm\sigma}\cdot{\bm\Psi}^*(\bar
P)}\frac{1+e{\bm\sigma}\cdot\hat{\textbf{p}}}{2}\\
S_{23}=\lambda_2{\bm\sigma}\cdot{\bm\Psi}(P)\sum_{e\pm}\frac{1}{\nu^2+(p+e\mu)^2+\lambda^2_2{\bm\sigma}\cdot{\bm\Psi}^*(\bar
P){\bm\sigma}\cdot{\bm\Psi}(
P)}\frac{1+e{\bm\sigma}\cdot\hat{\textbf{p}}}{2}\\
S_{32}=-\lambda_2{\bm\sigma}\cdot{\bm\Psi}^*(-\bar
P)\sum_{e\pm}\frac{1}{\nu^2+(p+e\mu)^2+\lambda^2_2{\bm\sigma}\cdot{\bm\Psi}(-
P){\bm\sigma}\cdot{\bm\Psi}^*(-\bar
P)}\frac{1+e{\bm\sigma}\cdot\hat{\textbf{p}}}{2}\\
S_{34}=-\sum_{e\pm}\frac{i\nu+e(p-e\mu)}{\nu^2+(p-e\mu)^2+\lambda^2_2{\bm\sigma}\cdot{\bm\Psi}^*(-\bar
P){\bm\sigma}\cdot{\bm\Psi}(-
P)}\frac{1+e{\bm\sigma}\cdot\hat{\textbf{p}}}{2}\\
S_{41}=\lambda_2{\bm\sigma}\cdot{\bm\Psi}^*(\bar
P)\sum_{e\pm}\frac{1}{\nu^2+(p-e\mu)^2+\lambda^2_2{\bm\sigma}\cdot{\bm\Psi}(
P){\bm\sigma}\cdot{\bm\Psi}^*(\bar
P)}\frac{1+e{\bm\sigma}\cdot\hat{\textbf{p}}}{2}\\
S_{43}=-\sum_{e\pm}\frac{i\nu+e(p+e\mu)}{\nu^2+(p+e\mu)^2+\lambda^2_2{\bm\sigma}\cdot{\bm\Psi}^*(\bar
P){\bm\sigma}\cdot{\bm\Psi}(
P)}\frac{1+e{\bm\sigma}\cdot\hat{\textbf{p}}}{2}
\end{eqnarray}\label{A5}
\end{subequations}

In order to get rid of the Pauli matrices in the denominators of
the propagator in Eq.(\ref{A5}), we expanded $\bm\Psi$ in terms of
the two circular polarization basis as Eq.(\ref{circular}). We
have
\begin{eqnarray}
\nonumber{\bm\sigma}\cdot{\bm\Psi}(P){\bm\sigma}\cdot{\bm\Psi}^*(\bar
P)&=&{\bm\Psi}(P)\cdot{\bm\Psi}^*(\bar
P)+i{\bm\sigma}\cdot[{\bm\Psi}(P)\times{\bm\Psi}^*(\bar P)]\\
&=&(1+{\bm\sigma}\cdot\hat{\bf p})\phi_+(P)\phi_+^*(\bar
P)+(1-{\bm\sigma}\cdot\hat{\bf p})\phi_-(P)\phi_-^*(\bar P)
\end{eqnarray}
Therefore, the entry Eqs.(\ref{A5d}) can be written as
\begin{eqnarray}
\nonumber
S_{21}=-\frac{i\nu-(p+\mu)}{\nu^2+(p+\mu)^2+\lambda_2^2[(1+{\bm\sigma}\cdot\hat{\bf
p})\phi_+(P)\phi_+^*(\bar P)+(1-{\bm\sigma}\cdot\hat{\bf
p})\phi_-(P)\phi_-^*(\bar P)]}\frac{1-{\bm\sigma}\cdot\hat{\bf
p}}{2}\\
-\frac{i\nu+(p-\mu)}{\nu^2+(p-\mu)^2+\lambda_2^2[(1+{\bm\sigma}\cdot\hat{\bf
p})\phi_+(P)\phi_+^*(\bar P)+(1-{\bm\sigma}\cdot\hat{\bf
p})\phi_-(P)\phi_-^*(\bar P)]}\frac{1+{\bm\sigma}\cdot\hat{\bf
p}}{2}
\end{eqnarray}
It can be shown that
\begin{equation}
f({\bm\sigma}\cdot\hat{\bf p})\frac{1\pm{\bm\sigma}\cdot\hat{\bf
p}}{2}=f(\pm1)\frac{1\pm{\bm\sigma}\cdot\hat{\bf p}}{2}
\end{equation}
with $f({\bm\sigma}\cdot\hat{\bf p})$ is an arbitrary function of
${\bm\sigma}\cdot\hat{\bf p}$ as long as $f(\pm1)$ is
well-defined. Using this identity, we find
\begin{equation}
S_{21}=-\frac{i\nu-(p+\mu)}{\nu^2+(p+\mu)^2+2\lambda_2^2\phi_-(P)\phi_-^*(\bar
P)}\frac{1-{\bm\sigma}\cdot\hat{\bf
p}}{2}-\frac{i\nu+(p-\mu)}{\nu^2+(p-\mu)^2+2\lambda_2^2\phi_+(P)\phi_+^*(\bar
P)}\frac{1+{\bm\sigma}\cdot\hat{\bf p}}{2}\label{A4}
\end{equation}
The other entries of the quark propagator can be simplified along
the same lines. Furthermore the entries with $\bm\Psi$ or
$\bm\Psi^*$ in the numerators can be reduced following the
identity (\ref{39}). The final expression of the quark propagator
is presented at the end of Sec.II and is used to calculate the
condensation energy density in Eq.(\ref{28}).


\section{Vector spherical harmonics and addition formula}

To each momentum $\textbf{p}$, one defines two transverse spherical
harmonics \cite{jackson}:
\begin{equation}
{\bf Y}_{jm}^{(1)}(\hat{\bf p})={\bf X}_{jm}(\hat{\bf
p})=\frac{1}{\sqrt{j(j+1)}}{\bf L}Y_{jm}(\hat{\bf p})
\end{equation}
and
\begin{equation}
{\bf Y}_{jm}^{(2)}(\hat{\bf p})=\hat{\bf p}\times{\bf X}_{jm}(\hat
{\bf p})
\end{equation}
where the angular momentum operator ${\bf L}=-i{\bf p}\times{\bf\nabla}_{\bf p}$
with ${\bf\nabla}_{\bf p}$ the gradient operator with respect to ${\bf p}$.
It follows from the relation bewteen the scalar spherical harmonics and the
Wigner D-function that
\begin{equation}
{\bf X}_{jm}(\hat{\bf p})=\sqrt{\frac{2j+1}{4\pi j(j+1)}} <jm|{\bf
J}e^{-iJ_z\varphi}e^{-iJ_y\theta}|j0>^{*}
\end{equation}
It is straightforward to show that
\begin{equation}
\hat{\bf e}_{\pm}^*(\hat{\bf p})\cdot{\bf
J}e^{-iJ_z\varphi}e^{-iJ_y\theta}
=\frac{1}{\sqrt{2}}e^{-iJ_z\varphi}e^{-iJ_y\theta}J_{\pm}
\end{equation}
with $J_{\pm}=J_x\pm iJ_y$ the raising and lowering operators. Therefore
\begin{equation}
\hat{\bf e}_{\pm}^*(\hat{\bf p})\cdot {\bf X}_{jm}(\hat{\bf p})
=\sqrt{\frac{2j+1}{8\pi}}D_{m\pm1}^{j*}(\varphi,\theta,0)
\end{equation}
and we have
\begin{equation}
{\bf Y}_{jm}^{(1)}(\hat{\bf p})={\bf X}_{jm}(\hat{\bf
p})=\sqrt{\frac{2j+1}{8\pi}}[D_{m1}^{j^*}(\varphi,\theta,0)\hat{\bf
e}_+(\hat{\bf p})+D_{m-1}^{j^*}(\varphi,\theta,0)\hat{\bf
e}_-(\hat{\bf p})]\label{A6}
\end{equation}
and
\begin{equation}
{\bf Y}_{jm}^{(2)}(\hat{\bf p})=\hat{\bf p}\times{\bf
X}_{jm}(\hat{\bf p})
=-i\sqrt{\frac{2j+1}{8\pi}}[D_{m1}^{j^*}(\varphi,\theta,0)\hat{\bf
e}_+(\hat{\bf p})-D_{m-1}^{j^*}(\varphi,\theta,0)\hat{\bf
e}_-(\hat{\bf p})]\label{A7}
\end{equation}
Both ${\bf Y}_{jm}^{(1)}(\hat{\bf p})$ and ${\bf
Y}_{jm}^{(2)}(\hat{\bf p})$ are regular at poles $\theta=0$ and
$\theta=\pi$, so are $D_{m1}^{j^*}(\varphi,\theta,0)\hat{\bf
e}_+(\hat{\bf p})$ and $D_{m-1}^{j^*}(\varphi,\theta,0)\hat{\bf
e}_-(\hat{\bf p})$.

In order to obtain the addition formula for Wigner D-function, we
need to know the Euler angles of the product of two rotations with
Euler angles $(\varphi, \theta, 0)$ and
$(\varphi^{\prime},\theta^{\prime},0)$. The composition rule of
the transformation parameters of a Lie group is independent of the
representations. For the rotation group, the $J=1/2$
representation is the most convenient one to extract the
composition rule. For $J=1/2$, we have
\begin{equation}
D^{\frac{1}{2}}(\alpha, \beta,
\gamma)=e^{-\frac{i}{2}\sigma_3\alpha}e^{-\frac{i}{2}\sigma_2\beta}e^{-\frac{i}{2}\sigma_3\gamma}\label{1/2rep}
\end{equation}
with $\sigma_{2,3}$ the Pauli matrices. We denote the combined
rotation
\begin{equation}
D^{{\frac{1}{2}}^{\dag}}(\varphi,\theta,0)D^{\frac{1}{2}}(\varphi^{\prime},\theta^{\prime
},0)=D^{\frac{1}{2}}(\alpha, \beta, \gamma)
\end{equation}
with $\alpha, \beta, \gamma$ the Euler angles of the combined
rotation. Using Eq.(\ref{1/2rep}), we obtain
\begin{eqnarray}
&&\nonumber\left( \begin{array}{cc}
\cos\frac{\beta}{2}e^{-\frac{i}{2}(\alpha+\gamma)} &
-\sin\frac{\beta}{2}e^{-\frac{i}{2}(\alpha-\gamma)}\\
\sin\frac{\beta}{2}e^{\frac{i}{2}(\alpha-\gamma)} &
\cos\frac{\beta}{2}e^{\frac{i}{2}(\alpha+\gamma)}
\end{array}
\right)\\
&=&\left( \begin{array}{cc}
\cos\frac{\theta}{2}\cos\frac{\theta^{\prime}}{2}e^{\frac{i}{2}(\varphi-\varphi^{\prime})}+\sin\frac{\theta}{2}\sin\frac{\theta^{\prime}}{2}e^{-\frac{i}{2}(\varphi-\varphi^{\prime})}
&
-\cos\frac{\theta}{2}\sin\frac{\theta^{\prime}}{2}e^{\frac{i}{2}(\varphi-\varphi^{\prime})}+\sin\frac{\theta}{2}\cos\frac{\theta^{\prime}}{2}e^{-\frac{i}{2}(\varphi-\varphi^{\prime})}\\
-\sin\frac{\theta}{2}\cos\frac{\theta^{\prime}}{2}e^{\frac{i}{2}(\varphi-\varphi^{\prime})}+\cos\frac{\theta}{2}\sin\frac{\theta^{\prime}}{2}e^{-\frac{i}{2}(\varphi-\varphi^{\prime})}
&
\sin\frac{\theta}{2}\sin\frac{\theta^{\prime}}{2}e^{\frac{i}{2}(\varphi-\varphi^{\prime})}+\cos\frac{\theta}{2}\cos\frac{\theta^{\prime}}{2}e^{-\frac{i}{2}(\varphi-\varphi^{\prime})}
\end{array}
\right)
\end{eqnarray}
The equivalence of each block on both sides gives rise to
\begin{equation}
\cos\beta=\cos\theta\cos\theta^{\prime}+\sin\theta\sin\theta^{\prime}\cos(\varphi-\varphi^{\prime})=\hat{\bf
p}\cdot\hat{\bf p}^{\prime}
\end{equation}
and
\begin{subequations}
\begin{eqnarray}
e^{i(\alpha+\gamma)}\cos^2\frac{\beta}{2}=\hat{\bf e}_+(\hat{\bf p})\cdot\hat{\bf e}_-(\hat{\bf p}^{\prime})\\
e^{i(\alpha-\gamma)}\sin^2\frac{\beta}{2}=-\hat{\bf e}_+(\hat{\bf p})\cdot\hat{\bf e}_+(\hat{\bf p}^{\prime})\\
e^{-i(\alpha-\gamma)}\sin^2\frac{\beta}{2}=-\hat{\bf e}_-(\hat{\bf
p})\cdot\hat{\bf e}_-(\hat{\bf p}^{\prime})\\
e^{-i(\alpha+\gamma)}\cos^2\frac{\beta}{2}=\hat{\bf e}_-(\hat{\bf
p})\cdot\hat{\bf e}_+(\hat{\bf p}^{\prime})
\end{eqnarray}
\end{subequations}
These equations fix on the Euler angles of the combined rotation.
Therefore, the composition rule of two rotations for
representation of arbitrary J reads
\begin{equation}
\sum_mD_{m,1}^{J^*}(\varphi,\theta,0)D_{m,1}^J(\varphi^{\prime},\theta^{\prime},0)=D_{11}^J(\alpha,\beta,\gamma)=
\frac{d_{11}^J(\beta)\cos^2\frac{\beta}{2}}{\hat{\bf e}_+(\hat{\bf
p})\cdot\hat{\bf e}_-(\hat{\bf p}^{\prime})}
\end{equation}

\end{document}